\documentclass[11pt,fleqn]{article}

\usepackage{a4}
\usepackage{amstext}
\usepackage{amsfonts}
\usepackage{amssymb}
\usepackage{color}
\usepackage{epsfig}
\usepackage{mathtools}
\usepackage{ulem}

\newcommand{\gtapprox}{\raisebox{-0.5ex}{$\,\stackrel{>}{\scriptstyle\sim}\,$}}

\begin{document}


\begin{flushright}
DESY 11-177, ROM2F/2011/15, SFB/CPP-11-54
\end{flushright}

\begin{center}

{\huge \bf $\Lambda_{\overline{\textrm{MS}}}$ from the static potential for QCD with}

{\huge \bf $n_f = 2$ dynamical quark flavors}

\vspace{0.5cm}

\textbf{Karl Jansen} \\
DESY, Platanenallee 6, D-15738 Zeuthen, Germany \&\\ Dipartimento di Fisica, Universit$\grave{\rm a}$ di Roma ``Tor Vergata'' and INFN, Sezione di Roma 2,\\ Via della Ricerca Scientifica, I-00133 Roma, Italy 

\textbf{Felix Karbstein} \\
Helmholtz-Institut Jena, Helmholtzweg 4, D-07743 Jena, Germany \& Theoretisch-Physikalisches Institut, Friedrich-Schiller-Universit\"at Jena, Max-Wien-Platz 1, D-07743 Jena, Germany

\textbf{Attila Nagy} \\
Humboldt Universit\"at zu Berlin, Newtonstra{\ss}e 15, D-12489, Berlin, Germany

\textbf{Marc Wagner} \\
Goethe-Universit\"at Frankfurt am Main, Institut f\"ur Theoretische Physik, \\ Max-von-Laue-Stra{\ss}e 1, D-60438 Frankfurt am Main, Germany

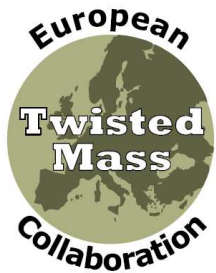
\vspace{0.7cm}

\begin{picture}(0,0)%
\includegraphics{Logo.pstex}%
\end{picture}%
\setlength{\unitlength}{4144sp}%
\begingroup\makeatletter\ifx\SetFigFont\undefined%
\gdef\SetFigFont#1#2#3#4#5{%
  \reset@font\fontsize{#1}{#2pt}%
  \fontfamily{#3}\fontseries{#4}\fontshape{#5}%
  \selectfont}%
\fi\endgroup%
\begin{picture}(1620,1620)(1,-781)
\end{picture}%

\vspace{0.4cm}

November 7, 2011

\end{center}

\vspace{0.1cm}

\begin{tabular*}{16cm}{l@{\extracolsep{\fill}}r} \hline \end{tabular*}

\vspace{-0.4cm}
\begin{center} \textbf{Abstract} \end{center}
\vspace{-0.4cm}

We determine $\Lambda_{\overline{\textrm{MS}}}$ for QCD with $n_f=2$ dynamical quark flavors by fitting the $Q \bar Q$ static potential known analytically in the perturbative regime up to terms of ${\cal O}(\alpha_s^4)$ and $\sim\alpha_s^4\ln\alpha_s$ to corresponding results obtained from lattice simulations. This has become possible, due to recent advances in both perturbative calculations, namely the determination and publication of the last missing contribution to the $Q \bar Q$ static potential at ${\cal O}(\alpha_s^4)$, and lattice simulations with $n_f=2$ dynamical quark flavors performed at the rather fine lattice spacing of $a \approx 0.042 \, \textrm{fm}$. Imposing conservative error estimates we obtain $\Lambda_{\overline{\textrm{MS}}} = 315(30) \, \textrm{MeV}$.

\begin{tabular*}{16cm}{l@{\extracolsep{\fill}}r} \hline \end{tabular*}

\thispagestyle{empty}


\newpage

\setcounter{page}{1}

\section{Introduction}

The last 15 years have seen substantial progress in the determination of the $Q\bar Q$ (quark-antiquark) static potential\footnote{In agreement with the prevalent notation, particularly in the field of lattice QCD, we use the terms {\it static potential} and {\it static energy} synonymously. Note, however, that sometimes a distinction is made and these terms refer to different quantities.} in the perturbative regime.
Whereas it had been known up to ${\cal O}(\alpha_s^2)$ for quite a long time \cite{Fischler:1977yf,Billoire:1979ih}, its next order contribution $\sim\alpha_s^3$ was determined only about 15 years ago, in the end of the 1990s \cite{Peter:1996ig,Peter:1997me,Schroder:1998vy}. It took another decade until all contributions at ${\cal O}(\alpha_s^4)$ were available in the $\overline{\rm MS}$-scheme \cite{Smirnov:2009fh,Anzai:2009tm,Chishtie:2001mf,Smirnov:2010zc,Smirnov:2008pn}.
A strict power-series expansion in $\alpha_s$, however, is not the whole story. Notably, already in the early days of quantum chromodynamics (QCD), it had been recognized \cite{Appelquist:1977es} that the static potential does not have a strict power-series expansion in $\alpha_s$. Beyond ${\cal O}(\alpha_s^3)$, also logarithmic contributions in $\alpha_s$ are induced. The leading one $\sim\alpha_s^4\ln\alpha_s$ has been determined explicitly, and in a systematic way by \cite{Brambilla:1999qa}. In summary, the static potential is now completely known up to ${\cal O}(\alpha_s^4)$ and $\sim\alpha_s^4\ln\alpha_s$.
On the other hand, the $Q\bar Q$ static potential is easily accessible in lattice QCD simulations (cf.\ e.g.\ \cite{Necco:2001xg,Donnellan:2010mx}), and has received continuous attention since the early days of lattice QCD. Having focused on the study of pure Yang-Mills theory without dynamical quark flavors, i.e.\ $n_f=0$, in the beginning, it has meanwhile become common to also account for light-flavor degrees of freedom. Moreover, the increase in available computing power has opened up the possibility to simulate on larger and larger lattices, thereby allowing for smaller and smaller lattice spacings.

Whereas the aforementioned perturbative calculations are conventionally performed in momentum space, lattice QCD naturally provides the static potential in position space. Due to the fact that QCD is asymptotically free, perturbation theory\footnote{We use the term {\it perturbation theory} in the sense, that the coupling $\alpha_s$ is small and can be dealt with perturbatively.} is viable at large momentum-transfers. After Fourier transform to position space, this allows for trustworthy insights at small $Q\bar Q$ separations. In contrast, lattice simulations at a given lattice spacing cannot resolve arbitrarily small separations.
The refinements on the lattice as well as the perturbative side have significantly reduced, and recently even bridged, the gap between the ranges of applicability of both approaches. This allows for a quantitative comparison of perturbative calculations with lattice simulations at $Q\bar Q$ separations, where both approaches are expected to be applicable.
While first exploratory attempts date back to the early 1990s \cite{Michael:1992nj}, 
recently a determination \cite{Brambilla:2010pp} of $r_0\Lambda_{\overline{\rm MS}}$ 
$n_f=0$ has been performed, resorting to the effective field theory of potential non-relativistic QCD (pNRQCD) \cite{Pineda:1997bj,Brambilla:1999xf} and the lattice results of \cite{Necco:2001xg}.
In this paper we focus on QCD with $n_f=2$ dynamical quark flavors and aim at the 
determination of $\Lambda_{\overline{\rm MS}}$ from the static potential.
In contrast to the aforementioned study at $n_f = 0$, we do not explicitly resort to pNRQCD, but directly confront the perturbatively determined $Q\bar Q$ static potential up to terms of ${\cal O}(\alpha_s^4)$ and $\sim\alpha_s^4\ln\alpha_s$ with lattice results.

Our paper is organized as follows. Whereas section~\ref{SEC599} outlines the determination of the $Q\bar Q$ static potential by means of lattice QCD, section~\ref{sec:pertth} focuses on its evaluation within perturbation theory. In section~\ref{sec:detofLambda} we describe, how $\Lambda_{\overline{\rm MS}}$ and the associated systematic and statistical errors are determined by fitting various orders of the perturbative expansion to lattice results of the $Q\bar Q$ static potential. We briefly compare our result for $\Lambda_{\overline{\rm MS}}$ at $n_f=2$ to previous determinations, and end with conclusions in section~\ref{sec:conclusions}.


\newpage

\section{\label{SEC599}The static potential from lattice QCD}

Although we compute the static potential for QCD with $n_f = 2$ dynamical quark flavors, 
employing standard methods from lattice gauge theory, we will give a description of our
lattice calculation below to make the paper self-contained. 


\subsection{\label{SEC852}Lattice setup}

We use $n_f = 2$ gauge link configurations generated by the European Twisted Mass Collaboration (ETMC)
\cite{Boucaud:2007uk,Boucaud:2008xu,Baron:2009wt}. 
The gauge action is tree-level Symanzik improved \cite{Weisz:1982zw},
\begin{eqnarray}
S_\mathrm{G}[U] \ \ = \ \ \frac{\beta}{6} \bigg(b_0 \sum_{x,\mu\neq\nu} \textrm{Tr}\Big(1 - P^{1 \times 1}(x;\mu,\nu)\Big) + b_1 \sum_{x,\mu\neq\nu} \textrm{Tr}\Big(1 - P^{1 \times 2}(x;\mu,\nu)\Big)\bigg)
\end{eqnarray}
with $b_0 = 1 - 8 b_1$ and $b_1 = -1/12$. The quark action is Wilson twisted mass (cf.\ \cite{Frezzotti:2000nk,Frezzotti:2003ni,Frezzotti:2004wz,Shindler:2007vp}),
\begin{eqnarray}
\label{EQN963} S_\mathrm{F}[\chi,\bar{\chi},U] \ \ = \ a^4 \sum_x \bar{\chi}(x) \Big(D_{\rm W} + i \mu_\mathrm{q} \gamma_5 \tau_3\Big) \chi(x) , 
\end{eqnarray}
with
\begin{eqnarray}
D_\mathrm{W} \ \ = \ \ \frac{1}{2} \Big(\gamma_\mu \Big(\nabla_\mu + \nabla^\ast_\mu\Big) - a \nabla^\ast_\mu \nabla_\mu\Big) + m_0 .
\end{eqnarray}
Here $a$ denotes the lattice spacing, $\nabla_\mu$ and $\nabla^\ast_\mu$ are the gauge covariant forward and backward derivatives, $m_0$ and $\mu_\mathrm{q}$ are the bare untwisted and twisted quark masses, respectively, $\tau_3$ is the third Pauli matrix acting in flavor space, and $\chi = (\chi^{(u)} , \chi^{(d)})$ represents the quark fields in the so-called twisted basis. The twist angle $\omega$ is given by $\tan\omega = \mu_\mathrm{R} / m_\mathrm{R}$, where $\mu_\mathrm{R}$ and $m_\mathrm{R}$ denote the renormalized twisted and untwisted quark masses. $\omega$ has been tuned to $\pi / 2$ by adjusting $m_0$ appropriately (cf.\ \cite{Boucaud:2008xu} for details). This ensures automatic $\mathcal{O}(a)$ improvement for many observables, including the static potential.

The ensembles of gauge link configurations considered here are collected in Table~\ref{TAB077}. We have data for four different values of the lattice spacing, a wide range of pion masses $m_\textrm{PS}$ and spacetime volumes $L^3 \times T$. Details on the generation of these gauge field configurations as well as on the computation and the analysis of standard quantities (e.g.\ lattice spacing or pion mass) can be found in \cite{Boucaud:2008xu,Baron:2009wt}. We also provide the number of gauge link configurations, used in the computation of the static potential, for each ensemble.




 












\begin{table}[htb]
\begin{center}
\begin{tabular}{|c|c|c|c|c|}
\hline
 & & & & \vspace{-0.40cm} \\
$\beta$ & $a$ in $\textrm{fm}$ & $(L/a)^3 \times T/a$ & $m_\textrm{PS}$ in $\textrm{MeV}$ & \# gauges \\
 & & & & \vspace{-0.40cm} \\
\hline
 & & & & \vspace{-0.40cm} \\
\hline
 & & & & \vspace{-0.40cm} \\
$3.90$ & $0.079(3)\phantom{00}$ & $24^3 \times 48$ & $340(13)$ & $168$ \\
 & & & & \vspace{-0.40cm} \\
\hline
 & & & & \vspace{-0.40cm} \\
$4.05$ & $0.063(2)\phantom{00}$ & $32^3 \times 64$ & $325(10)$ & $\phantom{0}71$ \\
       &                        &                  & $449(14)$ & $100$ \\
       &                        &                  & $517(16)$ & $\phantom{0}92$ \\
 & & & & \vspace{-0.40cm} \\
\hline
 & & & & \vspace{-0.40cm} \\
$4.20$ & $0.0514(8)\phantom{0}$ & $24^3 \times 48$ & $284(5)\phantom{0}$ & $123$ \\
       &                        & $48^3 \times 96$ &                     & $\phantom{0}46$ \\
 & & & & \vspace{-0.40cm} \\
\hline
 & & & & \vspace{-0.40cm} \\
$4.35$ & $0.0420(17)$           & $32^3 \times 64$ & $352(22)$ & $146$\vspace{-0.40cm} \\
 & & & & \\
\hline
\end{tabular}

\caption{\label{TAB077}Ensembles of gauge link configurations; the scale has been set via the pion mass and the pion decay constant, using chiral perturbation theory (cf.\ e.g.\ \cite{Boucaud:2008xu,Baron:2009wt}).}

\end{center}
\end{table}


\subsection{\label{SEC497}Computation of the static potential}

As usual, the static potential $V(r)$ at $Q\bar Q$ separation $r=|\vec{r}|$ is extracted from the exponential decay of Wilson loop averages $\langle W(r,t) \rangle$ with respect to their temporal extent $t$, while keeping their spatial extent $r$ constant. We consider Wilson loops formed by APE smeared spatial links ($N_\textrm{APE} = 60$, $\alpha_\textrm{APE} = 0.5$ for all our gauge link ensembles; cf.\ \cite{Jansen:2008si} for details), and ordinary, i.e.\ unsmeared temporal links.

Given nowadays typical lattice spacings of $\gtapprox 0.04 \, \textrm{fm}$, there is only a small range of $Q\bar Q$ separations, where the static potential $V(r)$ can be computed reliably by means of both lattice QCD and perturbation theory. For our finest lattice spacing ($\beta = 4.35$ ensemble) this range roughly amounts to $3a \ldots 5a$ (cf.\ section~\ref{SEC329} for a detailed discussion). Therefore, it is desirable to obtain a large number of lattice data points in this interval, i.e.\ a resolution significantly finer than the lattice spacing. To this end, we do not only consider ordinary on-axis, but also so-called off-axis Wilson loops, whose spatial sides are not exclusively oriented parallel to the $x$-, $y$- or $z$-axis, respectively. The off-axis sides are constructed by optimally approximating a straight line through a product of on-axis links and/or two- and three-dimensional diagonal links. In terms of on-axis links, two dimensional diagonal links are defined as
\begin{eqnarray}
 U(x;x+\hat\mu+\hat\nu) \ \ = \ \ P_{SU(3)}\Big(U_{\mu}(x) U_{\nu}(x+\hat{\mu}) + U_{\nu}(x) U_{\mu}(x+\hat{\nu})\Big),
\end{eqnarray}
with the projector $P_{_{SU(3)}}$, implementing a projection to $SU(3)$,
\begin{eqnarray}
P_{SU(3)}(U) \ \ = \ \  \frac{U'}{\det(U')^{1/3}} \quad , \quad U' \ \ = \ \ U \big(U^{\dagger} U \big)^{1/2} .
\end{eqnarray}
Three dimensional diagonal links are computed analogously,
\begin{multline}
U(x;x+\hat\mu+\hat\nu+\hat\rho) \ \ = \\
= \ \ P_{SU(3)}\Big(U_{\mu}(x) U_{\nu}(x+\hat{\mu}) U_{\rho}(x+\hat{\mu}+\hat{\nu}) + U_{\mu}(x) U_{\rho}(x+\hat{\mu}) U_{\nu}(x+\hat{\mu}+\hat{\rho}) + \dots\Big) .
\end{multline}
The optimal approximation of the straight line in terms of the above mentioned links is constructed efficiently by employing the Bresenham algorithm \cite{Bresenham:1965:we}, which has already been applied to the computation of the $Q\bar Q$ static potential in previous works (cf.\ e.g.\ \cite{Bolder:2000un}). Following this approach, we have computed all Wilson loop averages $\langle W(|\vec{r}|,t) \rangle$ with $\vec{r} = \vec{n} a$, $\vec{n} \in \mathbb{Z}^3$ and $|\vec{r}| \leq 10 a$. To improve the signal quality, we have averaged over loops, which are related by discrete translational and cubic rotational symmetry.

The $Q\bar Q$ static potential $V(r)$ at separation $r$ is obtained in a two-step procedure. First, an effective potential is computed,
\begin{eqnarray}
V^\textrm{(effective)}(r,t) \ \ = \ \ \frac{1}{a} \ln\bigg(\frac{\langle W(r,t) \rangle}{\langle W(r,t+a) \rangle}\bigg).
\end{eqnarray}
Second, the $t$-independent quantity $V(r)$ is obtained by performing an uncorrelated $\chi^2$ minimizing fit to $V^\textrm{(effective)}(r,t)$ in a suitable $t$ range. This range is chosen such that excited states are strongly suppressed, while statistical errors are still small.


Even though physical observables are automatically $\mathcal{O}(a)$ improved, as we are employing Wilson twisted mass lattice QCD at maximal twist (cf.\ section~\ref{SEC852}), we reduce lattice discretization errors even further. To this end, we use a method to improve the static potential, which is explained in \cite{Necco:2001xg,Necco:2003jf}. The improved potential is defined as follows,
\begin{eqnarray}
V^\textrm{(improved)}(r^\textrm{(improved)}) \ \ = \ \ V(r) ,
\end{eqnarray}
with $r^\textrm{(improved)}$ determined from
\begin{eqnarray}
\label{EQN612} \frac{1}{4 \pi r^\textrm{(improved)}} \ \ = \ \ \frac{1}{(2 \pi)^3} \int_{-\pi}^{+\pi} d^3k \frac{\prod_{j=1}^3 \cos(k_j r_j)}{4 \sum_{j=1}^3 \sin^2(k_j/2) + (4/3) \sum_{j=1}^3 \sin^4(k_j/2)}
\end{eqnarray}
(cf.\ \cite{Necco:2003jf}, section~3.2.2). This procedure assures that at tree level the lattice static potential computed with the tree-level Symanzik improved action is identical to the continuum static potential, i.e.\ $V^\textrm{(improved)}$ is a tree-level improved observable. $V^\textrm{(improved)}$ is the quantity to be confronted with perturbative calculations in section~\ref{sec:detofLambda}.

We have solved the integral (\ref{EQN612}) numerically via standard Monte Carlo sampling. Decomposing the integral and treating its singular part analytically, we were able to reach a precision, where statistical errors are negligible (cf.\ appendix~\ref{SEC411} for details).

%
%
%
%
%
%
%

\begin{figure}[htb]
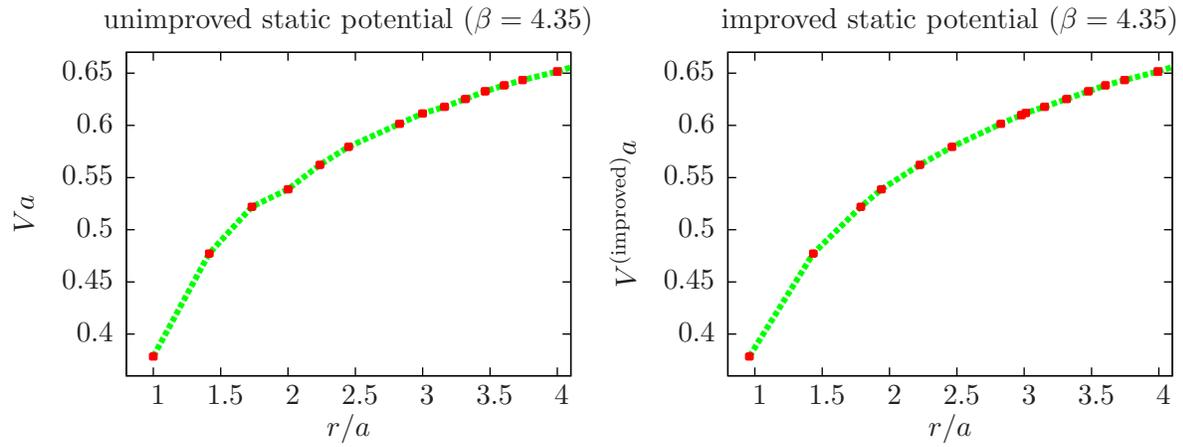

\begin{center}
\input{FIG001a.tex}\input{FIG001b.tex}
\caption{\label{FIG001}Comparison of the unimproved (left) and the improved (right) static potential for $\beta = 4.35$; the solid straight lines connecting the data points are drawn to guide the eye.}
\end{center}
\end{figure}

Figure~\ref{FIG001} compares the unimproved (left plot) and the improved (right plot) static potential for $\beta = 4.35$. In the unimproved static potential, discretization errors are clearly visible at small $Q\bar Q$ separations, where they amount to a strong violation of rotational invariance. On the other hand, the improved potential is rather smooth and no such violation is visible.

\newpage

$\quad$


\newpage

\section{\label{sec:pertth}The static potential in perturbation theory}

We now turn to the $Q\bar Q$ static potential in perturbation theory.
It is completely known up to terms of order $\alpha_s^4(\mu)$ and the contribution $\sim\alpha_s^4(\mu)\ln\alpha_s(\mu)$, where $\mu$ refers to an {\it a priori} arbitrary, but fixed momentum scale in the perturbative regime, i.e.\ $\mu\gg\Lambda_{\rm QCD}$. As we are interested in QCD, we limit our considerations to the color gauge group SU(3). Instead of referring to the corresponding literature only, we aim at keeping this paper rather self-contained. Therefore, we explain and explicitly provide all the formulae relevant for this work here.

As a matter of fact, most of the calculations in perturbative QCD are conveniently performed in momentum space. This is also true for the static potential. However, aiming at a comparison with lattice simulations naturally performed in position space, we are ultimately interested in a perturbative expression of the static potential in position space.

\subsection{\label{SEC499}The static potential in momentum space}

Defining \cite{Anzai:2010td}
\begin{equation}
 L\ \ \equiv \ \ L(\mu,p) \ \ = \ \ \ln\frac{\mu^2}{p^2},
\end{equation}
with $p=|\vec{p}|$, the static potential in momentum space can be written as
\begin{equation}
 \tilde V(p) \ \ = \ \ -C_F\frac{4\pi}{p^2}\,\tilde\alpha_{V}[\alpha_s(\mu),L(\mu,p)] , \label{eq:Vppert}
\end{equation}
where
\begin{multline}
 \tilde\alpha_{V}[\alpha_s(\mu),L(\mu,p)] \ \ = \ \ \alpha_s(\mu)\left\{1+\frac{\alpha_s(\mu)}{4\pi}\,P_1(L)
+\left(\frac{\alpha_s(\mu)}{4\pi}\right)^2\,P_2(L)\right. \\
\left.+\left(\frac{\alpha_s(\mu)}{4\pi}
\right)^3\Bigl[P_3(L)+a_{3 {\rm
ln}}\ln{\alpha_s}(\mu)\Bigr]+\ldots
\right\} . \label{eq:pert1}
\end{multline}
All the terms written explicitly in eq.~(\ref{eq:pert1}) are known. For SU(3) $C_F=4/3$.
The expansion coefficients $P_{n}(L)$ are polynomials in $L$ \cite{Anzai:2010td} and are stated below (cf.\ eqs.\ (\ref{eq:P1L}) to (\ref{eq:P2L})).

Up to ${\cal O}(\alpha_s^3(\mu))$ the static potential has a strict power-series expansion in $\alpha_s(\mu)$. Beyond this order the power-series expansion in $\alpha_s(\mu)$ breaks down \cite{Appelquist:1977es} and one also encounters logarithmic contributions in $\alpha_s(\mu)$.
The first such term is $\sim \alpha_s^4(\mu)\ln{\alpha_s(\mu)}$.
The contribution linear in $\alpha_s$ in eq.~(\ref{eq:pert1}) is the leading order (LO) expression of the static potential. Terms up to ${\cal O}(\alpha^2_s)$ correspond to next-to-leading order (NLO), and up to ${\cal O}(\alpha^3_s)$ to next-to-next-leading order (NNLO). Finally, terms up to ${\cal O}(\alpha^4_s)$ together with the $\alpha_s^4\ln{\alpha_s}$ contribution are referred to as NNNLO in the following.

Let us emphasize that, whereas we wrote the static potential by means of a perturbative expansion in terms of the coupling $\alpha_s(\mu)$ at a fixed momentum-scale $\mu$ in eqs.~(\ref{eq:Vppert}) and (\ref{eq:pert1}), the full 
static potential should not depend on any externally set scale. It should rather form a renormalization group (RG) invariant. This is true order by order in a perturbative expansion in $\alpha_s$ also.
Hence, $\tilde\alpha_{V}[\alpha_s,L]$, exhibiting an implicit $\mu$-dependence via both its arguments $\alpha_s$ and $L$, has to obey the following renormalization group equation \cite{Chishtie:2001mf},
\begin{eqnarray}
 \mu\frac{{\rm d}}{{\rm d}\mu}
\tilde\alpha_V[\alpha_s(\mu),L(\mu,p)] \ \ = \ \ 0\quad\quad\leftrightarrow\quad\quad\left(\frac{\partial
} { \partial L}+\frac{\alpha_s}{2}\beta[\alpha_s(\mu)]\frac{\partial}{\partial
\alpha_s}\right)\tilde\alpha_V[\alpha_s,L] \ \ = \ \ 0 . \label{dW=0}
\end{eqnarray}
Here $\beta[\alpha_s(\mu)]$ denotes the QCD $\beta$-function, characterizing the running of the coupling $\alpha_s$, defined as
\begin{equation}
\beta[\alpha_s(\mu)] \ \ \equiv \ \ \frac{\mu}{\alpha_s(\mu)}\frac{{\rm d}\alpha_s(\mu)}{{\rm
d}\mu} . \label{betafunction}
\end{equation}
It has the following power-series expansion in $\alpha_s(\mu)$,
\begin{equation}
\beta[\alpha_s(\mu)] \ \ = \ \ -\frac{\alpha_s(\mu)}{2\pi}\sum_{n=0}^{\infty}\left(\frac{\alpha_s(\mu)}{4\pi}
\right)^n\beta_n, \label{eq:betaseries}
\end{equation}
with the expansion coefficients $\beta_n$ known up to $n=3$, i.e.\ to $4$-loop order.
Whereas $\beta_0$ and $\beta_1$ are independent of the renormalization scheme,
$\beta_3$ and $\beta_4$ are scheme-dependent. They have been
determined for arbitrary compact semi-simple Lie groups in the $\overline{\rm
MS}$-scheme \cite{vanRitbergen:1997va}. For SU(3) they read
\begin{eqnarray}
 & & \hspace{-0.7cm} \beta_0 \ \ = \ \ 11-\frac{2}{3}n_f , \\
 & & \hspace{-0.7cm} \beta_1 \ \ = \ \ 102-\frac{38}{3}n_f , \\
 & & \hspace{-0.7cm} \beta_2 \ \ = \ \ \frac{2857}{2}-\frac{5033}{18}n_f+\frac{325}{54}n_f^2, \label{eq:beta_1} \\
\nonumber & & \hspace{-0.7cm} \beta_3 \ \ = \ \ \left(\frac{149753}{6}+3564\zeta(3)\right)-\left(\frac{1078361}{162}+\frac{6508}{27}\zeta(3)\right)n_f 
+\left(\frac{50065}{162}+\frac{6472}{81}\zeta(3)\right)n_f^2
\\
 & & \hspace{0.675cm} {}+\frac{1093}{729} n_f^3 \label{eq:beta_2}
\end{eqnarray}
with $n_f$ denoting the number of massless, dynamical quark flavors.

Eq.~(\ref{dW=0}) constrains the polynomials $P_n(L)$ in eq.~(\ref{eq:pert1}) to be of the following form \cite{Chishtie:2001mf},
\begin{eqnarray}
P_1(L) &=& a_1+\beta_0 L\,, \label{eq:P1L}\\
P_2(L) &=& a_2+(2a_1\beta_0+\beta_1)L+\beta_0^2L^2 \,, \\
P_3(L) &=& a_3+
(3a_2\beta_0+2a_1\beta_1+\beta_2)L+\beta_0\left(3a_1\beta_0+\frac{5}{2}
\beta_1\right)L^2+\beta_0^3L^3 \label{eq:P2L}\,.
\end{eqnarray}
Prefactors of $\alpha_s^{n+1}(\mu)$ contain terms up to order $L^n$ ($n\in\mathbb{N}$). Hence, in the perturbative limit considered here the expansion is not only in $\alpha_s(\mu)$, but also in powers of $L$.
For eq.~(\ref{eq:pert1}) to yield trustworthy results it is, therefore, not enough to require that $\alpha_s(\mu)$ is small. Besides $\mu,p\gg\Lambda_{\rm QCD}$, in general one has also to ensure that $L \sim 1$, i.e.\ that $p$ and $\mu$ are of the same order. 

For fixed $n_f$, $a_n$ ($n=1,2,3$) and $a_{3{\rm ln}}$ are constants. Their values cannot be extracted from eq.~(\ref{eq:pert1}), but necessitate an explicit determination.
The constants $a_1$ \cite{Fischler:1977yf,Billoire:1979ih} and $a_2$ \cite{Peter:1996ig,Peter:1997me,Schroder:1998vy} are known analytically. For gauge group SU(3) and in the $\overline{\rm MS}$-schema, they read
\begin{eqnarray}
 a_1&=&\frac{31}{3}-\frac{10}{9}n_f\,, \\
 a_2&=&\left(\frac{4343}{18}+36\pi^2-\frac{9}{4}\pi^4+66
\zeta(3)\right)-\left(\frac{1229}{27}+\frac{52}{3}\zeta(3)\right)
n_f +\frac{100}{81}n_f^2\,.
\end{eqnarray}
The logarithmic contribution in eq.~(\ref{eq:pert1}) is conveniently determined in position space \cite{Brambilla:1999qa,Brambilla:2006wp}.
Within the effective field theory of potential non-relativistic QCD (pNRQCD) \cite{Pineda:1997bj,Brambilla:1999xf} it naturally arises by taking into account the ultrasoft (US) contribution to the static potential. However, it can also be determined more directly by resumming a certain class of diagrams (Coulomb ladder diagrams) in perturbation theory without resorting to pNRQCD \cite{Appelquist:1977es,Anzai:2010td}.
As it also contributes to $a_3$, we postpone the specification of the explicit expressions for $a_3$ and $a_{3{\rm ln}}$, until having performed the transition to position space.

\subsection{\label{sec:posspace} The static potential in position space}

The static potential in position space is determined by means of a Fourier transform,
\begin{equation}
 V(r)\ \ =\ \ \int\frac{{d^3 p}}{(2\pi)^3}\,{\rm e}^{i\vec{p}\cdot\vec{r}}\,\tilde{V}(p)\,.
\label{E0}
\end{equation}
In section~\ref{SEC499}, we argued that, for the perturbative expansion of the potential to yield trustworthy results in momentum-space, $p$ and $\mu$ should be of the same order.
Conversely, in eq.~(\ref{E0}) the integral is over the full momentum regime.
This seems to be a contradiction at first glance only.
Let us first motivate that an explicit restriction to the perturbative momentum regime, $p\gtrsim\Lambda_{\rm QCD}$, would basically reproduce eq.~(\ref{E0}) up to a $r$-independent global shift term $V_{\rm shift}$, as
\begin{align}
\int\frac{{d^3 p}}{(2\pi)^3}\,{\rm e}^{i\vec{p}\cdot\vec{r}}\,\tilde V(p)\,\Theta(|\vec{p}|-\Lambda_{\rm QCD}) \nonumber\\
&\hspace*{-3cm}=\ \ \int\frac{{d^3 p}}{(2\pi)^3}\,{\rm e}^{i\vec{p}\cdot\vec{r}}\,
\tilde V(p) -\int\frac{{d^3 p}}{(2\pi)^3}\,{\rm e}^{i\vec{p}\cdot\vec{r}}\,\tilde V(p)\,\Theta(\Lambda_{\rm QCD}-|\vec{p}|)\,,
\intertext{and, because in the perturbative regime $\Lambda_{\rm QCD}\ll\frac{1}{r}$,}
&\hspace*{-3cm}\approx\ \ \int\frac{{d^3 p}}{(2\pi)^3}\,{\rm e}^{i\vec{p}\cdot\vec{r}}\,
\tilde V(p) -\int\frac{{d^3 p}}{(2\pi)^3}\,\tilde V(p)\,\Theta(\Lambda_{\rm QCD}-|\vec{p}|)\,\nonumber\\
&\hspace*{-3cm}=\ \ \int\frac{{d^3 p}}{(2\pi)^3}\,{\rm e}^{i\vec{p}\cdot\vec{r}}\,
\tilde V(p) - V_{\rm shift}\,.
\end{align}
As the lattice potential is known up to a global shift only, for our purposes the explicit expression for $V_{\rm shift}$ is not relevant. Second,
as the coefficient of a given power of $\alpha_s^n(\mu)$, $n\in\mathbb{N}$, encodes the full momentum dependence at this order, the Fourier transform can be performed order by order in $\alpha_s(\mu)$, irrespective of the relative relation between $p$ and $\mu$.
By inspection of its result, cf. eq.~(\ref{E02}), one finds that to prevent the respective logarithms from becoming large, one now has to require that $1/r$ does not deviate much from $\mu$. 
Hence, the resulting potential in position space $V(r)$ can be considered as trustworthy in the regime, where $\mu,\frac{1}{r}\gg\Lambda_{\rm QCD}$, and moreover $\mu$ and $1/r$ are of the same order. 

Introducing \cite{Anzai:2010td}
\begin{eqnarray}
 L' \ \ \equiv \ \ L'(\mu,r) \ \ = \ \ \ln(\mu^2r^2)+2\gamma_E, \label{eq:Lstrich}
\end{eqnarray}
where $\gamma_E$ denotes the Euler-Mascheroni constant, eq.~(\ref{E0}) immediately results in\begin{multline}
V(r) \ \ = \ \ -C_F\frac{
\alpha_s(\mu)}{r}\left\{1+\frac{\alpha_s(\mu)}{4\pi}\tilde{P}_1(L')+\left(\frac { \alpha_s(\mu)}{4\pi} \right)^2\tilde{P}
_2(L')\right. \\
\left.+\left(\frac{\alpha_s(\mu)}{4\pi}\right)^3\left[\tilde{P}_3(L')+a_{3{\rm
ln}}\ln\alpha_s(\mu)\right]+\ldots \right\}
\label{E02}
\end{multline}
with \cite{Anzai:2010td}
\begin{eqnarray}
\tilde{P}_1(L')&=&a_1+\beta_0 L', \\
\tilde{P}_2(L')&=&a_2+(2a_1\beta_0+\beta_1)
L'+\beta_0^2\left(L'^2+\frac{\pi^2}{3}\right),  \\
\tilde{P}_3(L')&=&a_3
+(3a_2\beta_0+2a_1\beta_1+\beta_2)L'+\left(3a_1\beta_0^2+\frac{5}{2}
\beta_0\beta_1\right)\left(L'^2+\frac{\pi^2}{3}\right) \nonumber\\
&&{}+\beta_0^3\left[L'^3+\pi^2 L'+16\zeta(3)\right].
\end{eqnarray}
Finally, we explicitly specify the coefficients $a_3$ and $a_{3{\rm ln}}$.
A comparison with eq.~(21) of \cite{Anzai:2010td} yields
\begin{eqnarray}
 a_3+a_{3{\rm ln}}\ln(\alpha_s(\mu)) \ \ \equiv \ \ \bar{a}_3+\frac{16}{3}\pi^2C_A^3\left[
\ln\left(C_A\alpha_s(\mu)\right)+\gamma_E-\frac{5}{6}\right], \label{2}
\end{eqnarray}
where, for SU(3), $C_A=3$ and \cite{Smirnov:2010zc,Anzai:2010td}
\begin{eqnarray}
\bar{a}_3 \ \ = \ \ a_3^{(0)}+a_3^{(1)}n_f+a_3^{(2)}n_f^2+a_3^{(3)}n_f^3
\end{eqnarray}
with
\begin{eqnarray}
a_3^{(0)}&=&27c_1 +\frac{15}{16}c_2, \\
a_3^{(1)}&=&\frac{9}{2}c_3+\frac{5}{96}c_4-\frac{68993}{81} +\frac{16624}{27}\zeta(3)+\frac{160}{9}\zeta(5)
, \\
a_3^{(2)}&=&\frac{93631}{972}+\frac{16}{45}\pi^4+\frac{412}{9}\zeta(3),\\
a_3^{(3)}&=&-\frac{1000}{729}.
\end{eqnarray}
The coefficients $c_i$ ($i=1\ldots4$) are only known numerically. $c_1$ and $c_2$ have been determined independently both in \cite{Smirnov:2009fh} and in \cite{Anzai:2009tm}. We use the numerical values from \cite{Smirnov:2009fh}, who provide smaller statistical errors\footnote{The errors associated with $c_1$ to $c_4$ turn out to be negligible in the context of our $\Lambda_{\overline{\textrm{MS}}}$ determination; therefore, we will not discuss them any further.},
\begin{equation}
 c_1 \ \ = \ \ 502.24(1) \quad , \quad
 c_2 \ \ = \ \ -136.39(12).
\end{equation}
$c_3$ and $c_4$ are given by \cite{Smirnov:2008pn},
\begin{equation}
 c_3 \ \ = \ \ -709.717 \quad , \quad
 c_4 \ \ = \ \ -56.83(1).
\end{equation}
From eq.~(\ref{2}) it follows that
\begin{eqnarray}
 & & \hspace{-0.7cm} a_3 \ \ = \ \ \bar{a}_3+\frac{16}{3}\pi^2C_A^3\left[
\ln\left(C_A\right)+\gamma_E-\frac{5}{6}\right] \ \ = \ \ \bar{a}_3+144\pi^2\left[
\ln3+\gamma_E-\frac{5}{6}\right], \\
 & & \hspace{-0.7cm} a_{3 {\rm ln}} \ \ = \ \ \frac{16}{3}\pi^2C_A^3=144\pi^2.
\end{eqnarray}
Therewith, all the coefficients appearing in the perturbative expansion
of the static potential, eq.~(\ref{eq:pert1}) and (\ref{E02}), respectively, have been assembled.
Only the scale $\mu$ and in particular the functional dependence of $\alpha_s(\mu)$ on $\mu$ have not yet been specified.

In particular note that eqs.~(\ref{betafunction}) and (\ref{eq:betaseries}) imply that $\alpha_s(\mu)$ can be expressed as a function of $\alpha_s(\nu)$ and $\ln(\mu^2/\nu^2)$.
Making a power-series ansatz for $\alpha_s(\mu)$ and requiring this expression to fulfill eqs.~(\ref{betafunction}) and (\ref{eq:betaseries}) one obtains
\begin{multline}
 \alpha_s(\mu) \ \ = \ \ \alpha_s(\nu)\left[1-\frac{\alpha_s(\nu)}{4\pi}\beta_0\ln\frac{\mu^2}{\nu^2}+\left(\frac{\alpha_s(\nu)}{4\pi}\right)^2\left(\beta_0^2\ln\frac{\mu^2}{\nu^2}-\beta_1\right)\ln\frac{\mu^2}{\nu^2} \right. \\
\left.-\left(\frac{\alpha_s(\nu)}{4\pi}\right)^3\left(\beta_0^3\ln^2\frac{\mu^2}{\nu^2}-\frac{5}{2}\beta_0\beta_1\ln\frac{\mu^2}{\nu^2}+\beta_2\right)\ln\frac{\mu^2}{\nu^2} \right]+{\cal O}\left(\alpha_s^5(\nu)\right), \label{eq:alpha_numu}
\end{multline}
i.e.\ prefactors of $\alpha^{n+1}_s(\nu)$ ($n\in\mathbb{N}$) contain terms up to order $\ln^n(\mu^2/\nu^2)$.
The expansion in eq.~(\ref{eq:alpha_numu}) is valid in the perturbative regime, i.e.\ for momenta $\mu$ and $\nu$ such that $\alpha_s(\mu)$ and $\alpha_s(\nu)$ are small. Moreover, $\mu$ and $\nu$ have to be of the same order, not to spoil the expansion by inducing large logarithms.
Inserting eq.~(\ref{eq:alpha_numu}) in the expressions for the static potential in momentum and position space, respectively, and keeping terms up to ${\cal O}(\alpha_s^4)$ and $\sim\alpha^4_s(\nu)\ln\alpha_s(\nu)$, we exactly recover eqs.~(\ref{eq:Vppert}) and (\ref{E02}) with $\mu$ substituted by $\nu$.
This clearly confirms that in principle, i.e.\ from a purely theoretical viewpoint detached from any phenomenological considerations, the static potential does not explicitly depend on any fixed scale $\mu$, order by order in a perturbative expansion, as demanded and discussed in the context of eq.~(\ref{dW=0}) above. Truncating at a given order, nothing favors a fixed given scale $\mu$ as compared to another fixed scale $\nu$, fulfilling the above requirements.

\subsection{Towards phenomenological applications}

In this section our aim is to make contact with phenomenology. First we line out, how the coupling $\alpha_s(\mu)$ at the scale $\mu$ is conventionally fixed within the framework of perturbation theory, namely by introducing a generic momentum scale $\Lambda$. Then we discuss, how to choose $\mu$ appropriately.

\subsubsection{\label{sec:alphaLambda} The scale $\Lambda_{\overline{\rm MS}}$ and its relation to the coupling $\alpha_s(\mu)$}

Recall, that in order to obtain eq.~(\ref{eq:alpha_numu}), we solved eqs.~(\ref{betafunction}) and (\ref{eq:betaseries}) by invoking a power-series ansatz. The result only enabled us to express the coupling at the scale $\mu$ in terms of the coupling at the scale $\nu$, without allowing us to identify any natural reference scale. Contrarily, the explicit integration of eq.~(\ref{betafunction}) naturally results in the introduction of a reference scale $\Lambda$.
Implementing the conventional normalization \cite{Bardeen:1978yd,Furmanski:1981cw} and keeping terms beyond one-loop level in eqs.~(\ref{betafunction}) and (\ref{eq:betaseries}), the renormalization group invariant parameter $\Lambda$ in our conventions reads (cf.\ e.g.\ \cite{PDG,Capitani:1998mq})
\begin{equation}
 \Lambda \ \ \equiv \ \ \mu\left(\frac{\beta_0\alpha_s(\mu)}{4\pi}\right)^{-\frac{\beta_1}{2\beta_0^2}}\exp\left\{-\frac{2\pi}{\beta_0\alpha_s(\mu)}-\int_0^{2\sqrt{\pi\alpha_s(\mu)}}\frac{{\rm d}\alpha_s}{\alpha_s}\left[\frac{1}{\beta(\alpha_s)}+\frac{2\pi}{\beta_0\alpha_s}-\frac{\beta_1}{2\beta_0^2}\right]\right\}. \label{eq:Lambda2-4}
\end{equation}
If all coefficients beyond $\beta_0$ in eq.~(\ref{eq:betaseries}) are neglected, i.e.\ at one-loop accuracy,
\begin{equation}
 \Lambda \ \ \equiv \ \ \mu\exp\left\{-\frac{2\pi}{\beta_0\alpha_s(\mu)}\right\}. \label{eq:alpha_1l}
\end{equation}
Eqs.~(\ref{eq:Lambda2-4}) and (\ref{eq:alpha_1l}) allow for the evaluation of $\alpha_s(\mu)$ as a function of the dimensionless ratio $\mu/\Lambda$.
Hence fixing the scale $\Lambda$ in principle amounts to knowing $\alpha_s(\mu)$ at any momentum scale in the perturbative regime.
As the QCD $\beta$-function is known at four-loop order only (cf.\ eq.~(\ref{eq:betaseries})), at best $\alpha_s(\mu)$ can be related to $\mu/\Lambda$ at four-loop accuracy. To obtain $\alpha_s(\mu)$ as a function of $\mu/\Lambda$ at two, three and four-loop accuracy, which we denote by $\alpha^{n{\rm -loop}}_s(\mu)$, with $n=2,3,4$ referring to the loop accuracy,  we insert the respective expressions for $\beta(\alpha)$ and solve eq.~(\ref{eq:Lambda2-4}) numerically. Only for $\alpha^{1{\rm -loop}}_s(\mu)$ a simple exact analytical solution can be inferred from eq.~(\ref{eq:alpha_1l}).
The quantity $\Lambda$ is renormalization scheme-dependent. With the expansion coefficients from eqs.~(\ref{eq:beta_1}) and (\ref{eq:beta_2}) evaluated in the $\overline{\rm MS}$-scheme, $\Lambda\equiv\Lambda_{\overline{\rm MS}}$.

\subsubsection{\label{sec:setscale} Setting the scale $\mu$}

We have manifestly required in eq.~(\ref{dW=0}), and subsequently also shown in the context of eq.~(\ref{eq:alpha_numu}), that the static potential does not explicitly depend on any particular choice of the externally set scale $\mu$ order by order in $\alpha_s$. Hence, in the full ``{\it all-order} expression'' of the static potential the choice of the scale $\mu$ would be completely arbitrary and the result would not depend on this scale at all.

In this work, however, our aim is to confront the truncated expression, eq.~(\ref{E02}), with lattice results within a certain range of $Q\bar Q$ separations, $r_{\rm min}\leq r\leq r_{\rm max}$, where perturbation theory is assumed to be valid.
Whenever truncating the series and only taking into account a finite number of terms, results certainly depend on the particular choice of the scale $\mu$.
In a range, where perturbation theory is valid and the requirement to be of the same order as $1/r$ is fulfilled for two different scales $\mu$ and $\nu$ (cf. the discussion above eq.~(\ref{eq:Lstrich})), the dependence of eq.~(\ref{E02}) on $\mu$ and $\nu$, respectively, is expected to be rather weak. Moreover, it should become smaller, when increasing the order in the perturbative expansion, thereby signalizing a reasonable convergence of the perturbative approximation. We demonstrate in section~\ref{sec:detofLambda} that this is indeed the case.

Following this philosophy, we fix the scale $\mu$ at a value, which is of the same order as $1/r$ in the spatial interval $r_{\rm min}\leq r\leq r_{\rm max}$, where the comparison with lattice results is going to be performed. A convenient choice is $\mu\equiv 2/(r_{\rm min} + r_{\rm max})$, which will also be adopted in the following section.


\newpage

\section{\label{sec:detofLambda} Determination of $\Lambda_{\overline{\textrm{MS}}}$}

In this section, we determine $\Lambda_{\overline{\textrm{MS}}}$ in units of the lattice spacing, i.e.\ $a\Lambda_{\overline{\textrm{MS}}}$, by fitting perturbative expressions for the static potential (cf.\ section~\ref{sec:pertth}) to corresponding lattice results (cf.\ section~\ref{SEC599}). Using the values of the lattice spacing listed in Table~\ref{TAB077}, these results can easily be converted to physical units, e.g.\ $\textrm{MeV}$. Unless explicitly specified otherwise, the errors on $\Lambda_{\overline{\textrm{MS}}}$ do not account for the uncertainties associated with the lattice spacing. We include these errors at the end, when quoting our final result for $\Lambda_{\overline{\textrm{MS}}}$ (cf.\ eq.\ (\ref{EQN844})).


\subsection{\label{SEC559}Fitting procedures}

Let us for the moment assume that we have already specified the scale $\mu$. To determine $\Lambda_{\overline{\textrm{MS}}}$ we then apply two strategies:
\begin{itemize}
\item[(A)] First we determine $\alpha_s(\mu)$ by means of fitting the perturbative expression for the $Q\bar Q$ static potential in eq.\ (\ref{E02}) to the lattice data points. To test the convergence of the perturbative expression, and to obtain an estimate for the systematic error, we do this for the four different orders, i.e.\ LO, NLO, NNLO and NNNLO, introduced in section~\ref{SEC499}. In a second step we always identify $\alpha_s(\mu)\equiv\alpha_s^{4-{\rm loop}}(\mu)$, i.e.\ employ the QCD $\beta$-function to the best accuracy available, and solve eq.\ (\ref{eq:Lambda2-4}) for $\Lambda_{\overline{\textrm{MS}}}$ (cf. also section~\ref{sec:alphaLambda}).

\item[(B)] Instead of always identifying $\alpha_s(\mu)$ with $\alpha_s^{4-{\rm loop}}(\mu)$ in the perturbative expression for the static potential, we explicitly make use of the various expressions relating $\alpha_s(\mu)$ to $\Lambda_{\overline{\rm MS}}/\mu$ at different loop accuracies, i.e.\ $\alpha_s^{l-{\rm loop}}(\mu)$ with $l=1,...,4$.
The identification of $\alpha_s(\mu)$ with $\alpha_s^{l-{\rm loop}}(\mu)$ is varied, depending both on the power $n$ of a given contribution $\sim\alpha^{n}_s(\mu)$ to the static potential, and on the order in $\alpha_s(\mu)$, up to which the potential itself is expanded to, i.e.\ LO, NLO, NNLO and NNNLO.
At NNNLO the coupling $\alpha_s(\mu)$ in terms proportional to $\alpha^n_s(\mu)$ is identified with $\alpha_s^{l-{\rm loop}}(\mu)$, where $l\equiv5-n$, and the $\alpha_s(\mu)$ in the logarithm with $\alpha_s^{1-{\rm loop}}(\mu)$; at NNLO the identification works analogously, but $l\equiv4-n$, at NLO, $l\equiv 3-n$ and at LO, $l\equiv 2-n$.
The resulting expressions are functions of $r$ and $\Lambda_{\overline{\rm MS}}/\mu$. Consequently, $\Lambda_{\overline{\textrm{MS}}}$ can be determined by a fit.
\end{itemize}
Let us remark that both strategies (A) and (B) seem to be equally justified. There is no compelling reason favoring any of them. In particular both strategies ensure that at least all terms up to a given order in $\alpha_s(\mu)$ are consistently taken into account.

We perform these fits to various lattice results for the static potential, corresponding to the ensembles of gauge link configurations listed in Table~\ref{TAB077}. The fit is an uncorrelated $\chi^2$ minimizing two parameter fit of the perturbative expansion of the static potential $V(r)$ to the corresponding lattice results in the range $r_\textrm{min} \leq r \leq r_\textrm{max}$, where $r_\textrm{min}$, $r_\textrm{max}$ and $\mu$ are input parameters, discussed below. The fit parameters are $a V_0$ (a constant shift of the static potential) and $\alpha_s(\mu)$ (fitting procedure (A)) or $a \Lambda_{\overline{\textrm{MS}}}$ (fitting procedure (B)), respectively.


\subsection{Some general comments on our approach}

In passing, we shortly provide some general comments on our approach. This is intended to ease the reader in relating it to other works.

The first point to note, is that our approach is fully compatible with arguments emphasizing the necessity to account for renormalon contributions, in order to enhance the convergence behavior of the perturbative expansion. As outlined in detail in \cite{Pineda:2002se}, the expansion in terms of $\alpha_s(\mu)$, evaluated at the fixed momentum scale $\mu$, and the allowance for a constant shift of the static potential as independent fitting parameter at each order in the expansion of the $Q\bar Q$ static potential, as employed by us, take the renormalon effects fully into account \cite{Pineda:2002se}. 

Second, a lot of work has been invested to determine the logarithmic contributions to the static potential. Those at ${\cal O}(\alpha_s^5)$ have been determined explicitly in \cite{Brambilla:2006wp}.
Moreover, resorting to the renormalization group, the leading ultrasoft logarithms \cite{Pineda:2000gza} and next-to-leading ultrasoft logarithms \cite{Brambilla:2009bi} in the static potential could be resummed.
This can be particularly important, when aiming at a study of the short-distance behavior of the static potential.
In our work, we do not invoke the resummation of logarithms.
Aiming at a direct comparison and fitting of the perturbative expressions with data points from lattice QCD simulations, we are naturally in the large-distance sector of the perturbative regime, where $\alpha_s$ is not particularly small, but rather of the order of a few tenths. Consequently $\ln\alpha_s$ is of order $1$ and, hence, not expected to dominate the other (unknown) contributions at a given order in the expansion in $\alpha_s$.


\subsection{\label{SEC329}Input parameters to the fitting procedure and systematic errors of $\Lambda_{\overline{\textrm{MS}}}$ \\ arising from the perturbative side}


\subsubsection{Individual variation of input parameters}

In this section, we investigate the stability of $\Lambda_{\overline{\textrm{MS}}}$ with respect to the input parameters $r_\textrm{min}$, $r_\textrm{max}$ and the scale $\mu$, at which the coupling constant $\alpha_s(\mu)$ is defined, using our finest lattice spacing (ensemble with $\beta = 4.35$). To this end we vary the input parameters in the following intervals:
\begin{itemize}
\item $r_\textrm{min} = 2a \ldots 4a$: \\
lattice discretization errors in the static potential typically become small, when increasing the $Q\bar Q$ separation beyond $2a \ldots 3a$. Due to the fact that our gauge action is tree-level Symanzik improved, and the lattice static potential is further improved, as explained in section~\ref{SEC497}, we expect only marginal discretization errors for $r \geq 2a$.

\item $r_\textrm{max} = 4a \ldots 6a$ (corresponding to $r_\textrm{max} = 0.17 \, \textrm{fm} \ldots 0.25 \, \textrm{fm}$): \\
the NNNLO expression for the static potential is expected to be in good agreement with lattice data up to separations of around $0.25 \, \textrm{fm}$ \cite{Pineda:2002se,Brambilla:2009bi}.

\item $1/\mu = 3a \ldots 5a$: \\
following the discussion in section~\ref{sec:setscale}, it is natural to choose $1/\mu$ in the vicinity of the fitting interval $r_\textrm{min} \ldots r_\textrm{max}$ (cf. also \cite{Pineda:2002se}).
\end{itemize}
Below, we demonstrate that the fit results for $\Lambda_{\overline{\textrm{MS}}}$ are rather stable with respect to these parameter variations, i.e.\ we confirm that a meaningful and a rather precise matching of perturbation theory and lattice results seems to be possible.

\begin{figure}[htb]
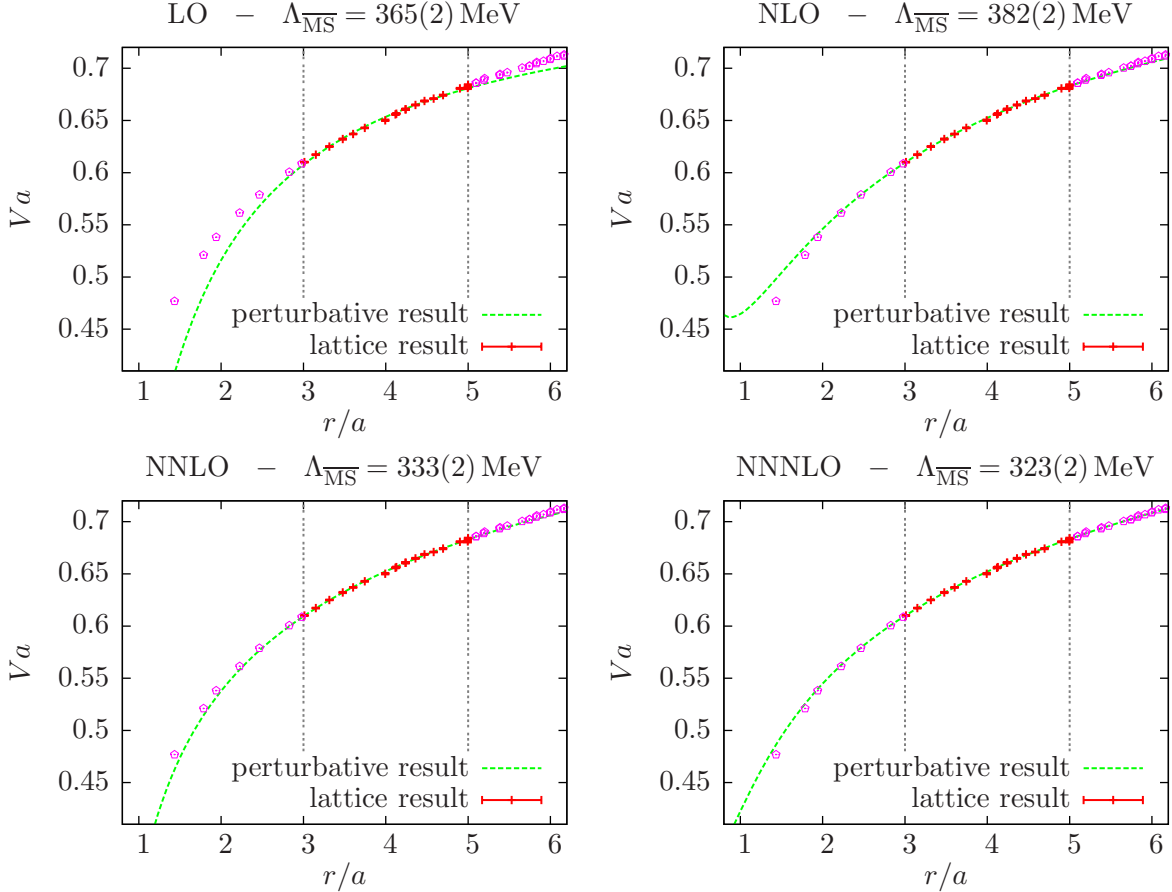

\begin{center}
\input{FIG002a.tex}\input{FIG002b.tex}
\input{FIG002c.tex}\input{FIG002d.tex}
\caption{\label{FIG002}Exemplary fits of the perturbative expressions at LO, NLO, NNLO and NNNLO to $\beta = 4.35$ lattice results. We employ fitting procedure (A), and set $r_\textrm{min} = 3a$, $r_\textrm{max} = 5a$, and $1/\mu = 4a$.}
\end{center}
\end{figure}

Exemplary fits of the perturbative orders, i.e.\ LO, NLO, NNLO and NNNLO, to lattice results are shown in Figure~\ref{FIG002}. Fitting procedure (A) has been used. The above mentioned input parameters have been chosen at the centers of their above defined ranges of variation, i.e.\ $r_\textrm{min} = 3a$, $r_\textrm{max} = 5a$ and $1/\mu = (r_\textrm{min} + r_\textrm{max}) / 2 = 4a$. The resulting values for $\Lambda_{\overline{\textrm{MS}}}$ values in $\textrm{MeV}$ are shown in the figure captions.

To understand, how the extracted $\Lambda_{\overline{\textrm{MS}}}$ depends on the input parameters $r_\textrm{min}$, $r_\textrm{max}$ and $1 / \mu$, we vary them individually in the following.

\begin{figure}[htb]
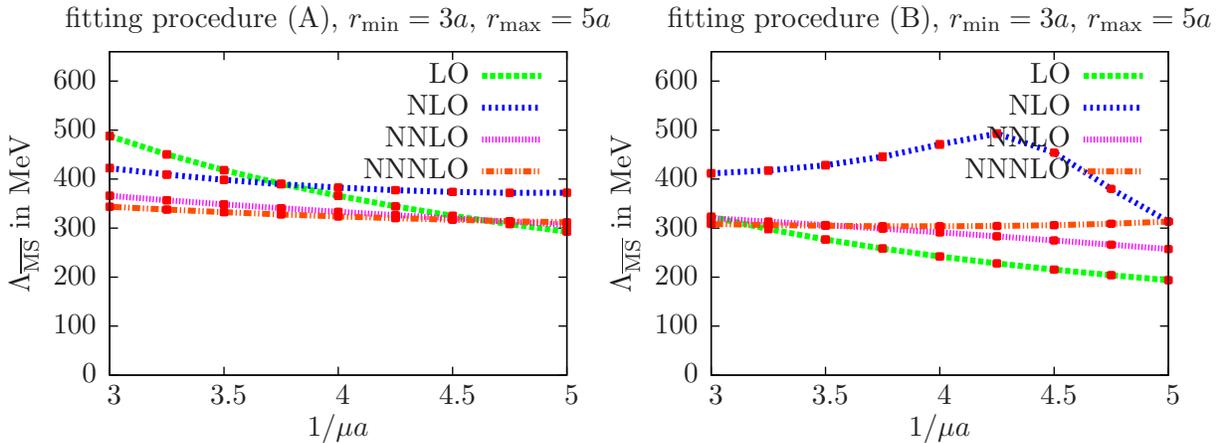

\begin{center}
\input{FIG003_mu__4_loop.tex}\input{FIG003_mu__1to4_loop.tex}
\caption{\label{FIG003a}Plot of $\Lambda_{\overline{\textrm{MS}}}$ as a function of $1/\mu$ at LO, NLO, NNLO, and NNNLO;
\textbf{left:} fitting procedure (A);
\textbf{right:} fitting procedure (B).
}
\end{center}
\end{figure}

In Figure~\ref{FIG003a} we vary the scale $1/\mu$ between $r_\textrm{min} = 3a$ and $r_\textrm{max} = 5a$. Moreover, we compare fitting procedure (A) and (B):
\begin{itemize}
\item In order to allow for a meaningful determination of $\Lambda_{\overline{\textrm{MS}}}$ it is important that the dependence of the fit results for $\Lambda_{\overline{\textrm{MS}}}$ on the scale $\mu$ is rather weak; this amounts to a plateaux-like behavior in the plot of $\Lambda_{\overline{\textrm{MS}}}$ as a function of $1/\mu$. Such behavior is clearly observable at NNNLO, and to some extent also at NNLO.

\item The expressions at NNNLO and NNLO yield similar results, indicating that higher orders do not contribute significantly.

\item The convergence of the four available orders, i.e.\ LO, NLO, NNLO and NNNLO, is better for fitting procedure (A) than for fitting procedure (B); this might be related to the fact that for (A) always the 4-loop expression for $\alpha_s(\mu)$ is used, i.e.\ that the fit formulae used for (B) contain less information.







\item Fitting procedure (A):
\begin{itemize}
\item Variation of $\Lambda_{\overline{\textrm{MS}}}$ (NNLO): $307 \, \textrm{MeV} \ldots 366 \, \textrm{MeV}$.

\item Variation of $\Lambda_{\overline{\textrm{MS}}}$ (NNNLO): $312 \, \textrm{MeV} \ldots 344 \, \textrm{MeV}$.
\end{itemize}







\item Fitting procedure (B):
\begin{itemize}
\item Variation of $\Lambda_{\overline{\textrm{MS}}}$ (NNLO): $257 \, \textrm{MeV} \ldots 320 \, \textrm{MeV}$.

\item Variation of $\Lambda_{\overline{\textrm{MS}}}$ (NNNLO): $304 \, \textrm{MeV} \ldots 313 \, \textrm{MeV}$.
\end{itemize}
\end{itemize}

\begin{figure}[htb]
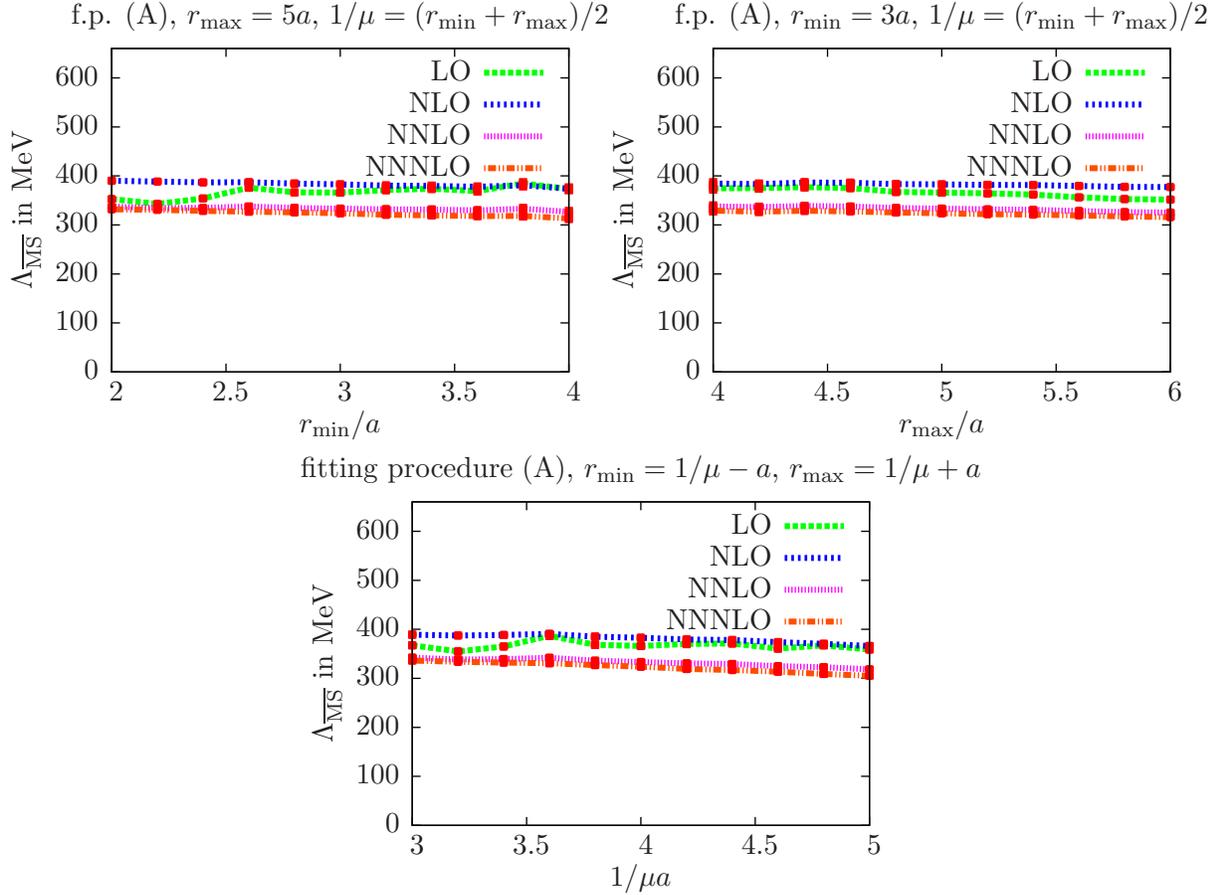

\begin{center}
\input{FIG003_r_min.tex}\input{FIG003_r_max.tex}
\input{FIG003_r_av.tex}
\caption{\label{FIG003b}$\Lambda_{\overline{\textrm{MS}}}$ at LO, NLO, NNLO, and NNNLO, fitting procedure (A);
\textbf{left:} as a function of $r_\textrm{min}$;
\textbf{right:} as a function of $r_\textrm{max}$; 
\textbf{bottom:} as a function of $1/\mu$.
}
\end{center}
\end{figure}

In Figure~\ref{FIG003b} we investigate the stability of the fit results for $\Lambda_{\overline{\textrm{MS}}}$ with respect to a variation of the fitting interval $r_\textrm{min} \ldots r_\textrm{max}$, applying fitting procedure (A); we vary both $r_\textrm{min}$ and $r_\textrm{max}$ separately as well as the center of the fitting interval, while keeping its extension $r_\textrm{max} - r_\textrm{min}$ fixed:
\begin{itemize}
\item Fit results for $\Lambda_{\overline{\textrm{MS}}}$ depend only weakly on the fitting range in particular for NNLO and NNNLO.






\item Varying $r_\textrm{min} = 2a \ldots 4a$, $r_\textrm{max} = 5a$, $1/\mu = (r_\textrm{min} + r_\textrm{max}) / 2$:
\begin{itemize}
\item Variation of $\Lambda_{\overline{\textrm{MS}}}$ (NNLO): $327 \, \textrm{MeV} \ldots 337 \, \textrm{MeV}$.

\item Variation of $\Lambda_{\overline{\textrm{MS}}}$ (NNNLO): $313 \, \textrm{MeV} \ldots 332 \, \textrm{MeV}$.
\end{itemize}






\item Varying $r_\textrm{max} = 4a \ldots 6a$, $r_\textrm{min} = 3a$, $1/\mu = (r_\textrm{min} + r_\textrm{max}) / 2$:
\begin{itemize}
\item Variation of $\Lambda_{\overline{\textrm{MS}}}$ (NNLO): $325 \, \textrm{MeV} \ldots 338 \, \textrm{MeV}$.

\item Variation of $\Lambda_{\overline{\textrm{MS}}}$ (NNNLO): $316 \, \textrm{MeV} \ldots 329 \, \textrm{MeV}$.
\end{itemize}






\item Varying $1/\mu = 3a \ldots 5a$, $r_\textrm{min} = 1/\mu - a$, $r_\textrm{max} = 1/\mu + a$:
\begin{itemize}
\item Variation of $\Lambda_{\overline{\textrm{MS}}}$ (NNLO): $318 \, \textrm{MeV} \ldots 342 \, \textrm{MeV}$.

\item Variation of $\Lambda_{\overline{\textrm{MS}}}$ (NNNLO): $305 \, \textrm{MeV} \ldots 336 \, \textrm{MeV}$.
\end{itemize}
\end{itemize}

\begin{figure}[htb]
\begin{center}
\input{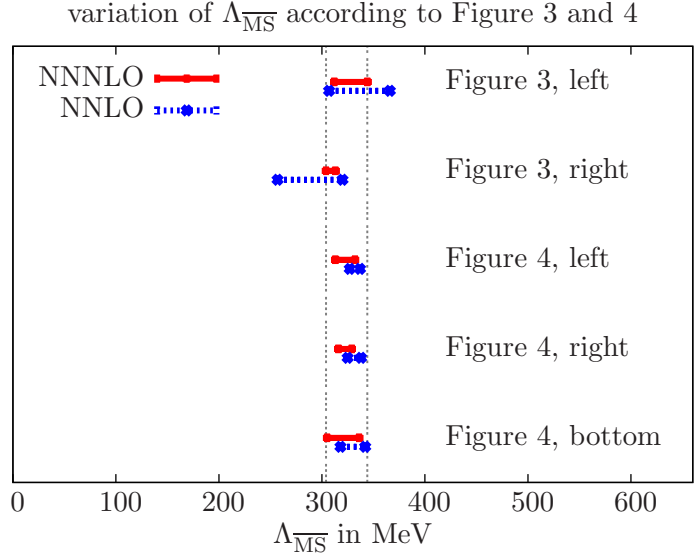}
\caption{\label{FIG003_error}Graphical summary of the variation of $\Lambda_{\overline{\textrm{MS}}}$, when altering the input parameters individually (cf.\ Figure~\ref{FIG003a} and Figure~\ref{FIG003b} and the corresponding text).}
\end{center}
\end{figure}

The variation of $\Lambda_{\overline{\textrm{MS}}}$ induced when varying the input parameters individually, as done in Figure~\ref{FIG003a} and Figure~\ref{FIG003b}, is summarized graphically in Figure~\ref{FIG003_error} both at NNLO (blue) and at (red). As systematic uncertainty one could quote, e.g.\ the whole range of values covered by the NNNLO variations,
\begin{eqnarray}
\label{EQN257} \Lambda_{\overline{\textrm{MS}}} \ \ = \ \ 304 \, \textrm{MeV} \ldots 344 \, \textrm{MeV} .
\end{eqnarray}
The sources of the systematic error investigated above might, however, be correlated. To be on the safe side one could add the errors in quadrature. Another more sophisticated method is discussed in section~\ref{SEC444}.

\subsubsection{\label{SEC444}Inclusion of correlations between different sources of the systematic error}

To account for possible correlations, we perform a large number of fits, with the input parameters chosen randomly and uniformly in the intervals specified above. As systematic error we then take the variance of the fit results.

We have performed 40,000 fits (sufficiently many that the statistical error of the variance is negligible):
\begin{itemize}
\item 10,000 NNLO fits, fitting procedure (A);

\item 10,000 NNLO fits, fitting procedure (B);

\item 10,000 NNNLO fits, fitting procedure (A);

\item 10,000 NNNLO fits, fitting procedure (B).
\end{itemize}
For these fits we randomly choose
\begin{itemize}
\item $r_\textrm{min} = 2a \ldots 4a$ and $r_\textrm{max} = 4a \ldots 6a$, imposing the constraint $r_\textrm{max} - r_\textrm{min} \geq a$;

\item $1 / \mu = r_\textrm{min} \ldots r_\textrm{max}$.
\end{itemize}
We obtain an average and a variance, i.e.\ a systematic error, of
\begin{eqnarray}
\label{EQN791} \Lambda_{\overline{\textrm{MS}}} \ \ = \ \ 315(26) \, \textrm{MeV} .
\end{eqnarray}
The error is slightly larger as compared to that stated in eq.~(\ref{EQN257}). However, both results perfectly agree within errors.

The fitting procedure introduces an additional statistical uncertainty of $\approx 2 \, \textrm{MeV}$, which is negligible, when added in quadrature.

Further systematic uncertainties, exclusively originating from the lattice static potential, are discussed in the next subsection.


\subsection{Systematic errors of $\Lambda_{\overline{\textrm{MS}}}$ associated with the lattice computation}

In the following we address lattice discretization errors and finite volume effects. Moreover, we investigate the dependence of the static potential on the light quark mass.


\subsubsection{\label{SEC330}Lattice discretization errors}

To get an idea about the order of magnitude of lattice discretization errors, we turn to the lightest quark mass available at each of the four values of the lattice spacing. As can be seen from Table~\ref{TAB077}, the corresponding pions are roughly of the same mass $\approx 284 \, \textrm{MeV} \ldots 352 \, \textrm{MeV}$. 
Keeping the parameters $r_\textrm{min}$ and $r_\textrm{max}$ in physical units approximately the same, we then extract values for $\Lambda_{\overline{\textrm{MS}}}$ in physical units, by performing fits (fitting procedure (A), NNNLO, $1/\mu = (r_\textrm{max} + r_\textrm{min}) / 2$),
\begin{itemize}
\item $\beta = 3.90$: \\
$r_\textrm{min} = 2.00 \times a$, $r_\textrm{max} = 3.50 \times a$
$\quad \rightarrow \quad$ $\Lambda_{\overline{\textrm{MS}}} = 317(12) \, \textrm{MeV}$.

\item $\beta = 4.05$: \\ 
$r_\textrm{min} = 2.53 \times a$, $r_\textrm{max} = 4.42 \times a$
$\quad \rightarrow \quad$ $\Lambda_{\overline{\textrm{MS}}} = 312(10) \, \textrm{MeV}$.

\item $\beta = 4.20$ ($L^3 \times T = 48^3 \times 96$): \\ 
$r_\textrm{min} = 3.14 \times a$, $r_\textrm{max} = 5.50 \times a$
$\quad \rightarrow \quad$ $\Lambda_{\overline{\textrm{MS}}} = 306(5) \, \textrm{MeV}$.

\item $\beta = 4.35$: \\ 
$r_\textrm{min} = 3.77 \times a$, $r_\textrm{max} = 6.60 \times a$
$\quad \rightarrow \quad$ $\Lambda_{\overline{\textrm{MS}}} = 304(12) \, \textrm{MeV}$.
\end{itemize}
%
%
%
Note, that the errors associated with the extracted $\Lambda_{\overline{\textrm{MS}}}$ values are rather large, because this time we have included the errors associated with the lattice spacings (cf.\ Table~\ref{TAB077}). This is essential, when comparing results obtained at different values of the lattice spacing.

\begin{figure}[htb]
\begin{center}
\input{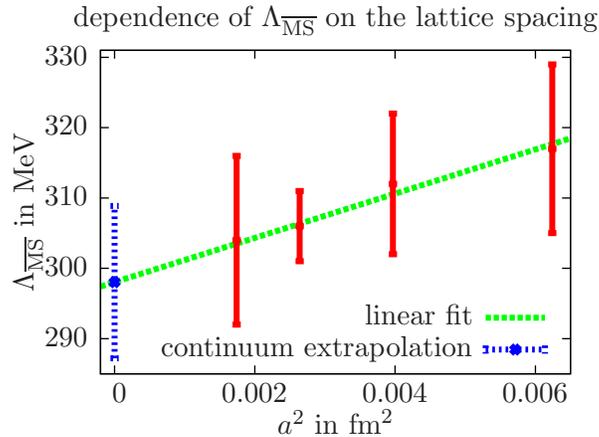}
\caption{\label{FIG004}Dependence of $\Lambda_{\overline{\textrm{MS}}}$ on the lattice spacing.}
\end{center}
\end{figure}

These results for $\Lambda_{\overline{\textrm{MS}}}$, as well as a continuum extrapolation assuming a dependence $\propto a^2$, are depicted in Figure~\ref{FIG004}. Within statistical errors, the result of the extrapolation, $\Lambda_{\overline{\textrm{MS}}} = 298(11) \, \textrm{MeV}$, is in agreement with the result at $\beta = 4.35$. Nevertheless, Figure~\ref{FIG004} indicates that there might be a slight downward tendency, when approaching the continuum limit. We are conservative and add an additional systematic error of $\pm 6 \, \textrm{MeV}$ to $\Lambda_{\overline{\textrm{MS}}}$, estimated by taking the difference between the central values of $\Lambda_{\overline{\textrm{MS}}}$ at our smallest lattice spacing and the continuum extrapolation.

\subsubsection{Finite volume effects}

We investigate finite volume effects by considering the two spacetime volumes at $\beta = 4.20$, which differ by a factor of $16$. Again we extract $\Lambda_{\overline{\textrm{MS}}}$ by performing fits with identical parameters $r_\textrm{min} = 3.14 \times a$ and $r_\textrm{max} = 5.50 \times a$:
\begin{itemize}
\item $\beta = 4.20$ ($L^3 \times T = 24^3 \times 48$): \quad
$\rightarrow$\quad $\Lambda_{\overline{\textrm{MS}}} = 305(1) \, \textrm{MeV}$.

\item $\beta = 4.20$ ($L^3 \times T = 48^3 \times 96$): \quad
$\rightarrow$\quad $\Lambda_{\overline{\textrm{MS}}} = 306(1) \, \textrm{MeV}$.
\end{itemize}
Within tiny statistical errors (compared to the systematic uncertainty of $\pm 26 \, \textrm{MeV}$ determined in section~\ref{SEC329}) $\Lambda_{\overline{\textrm{MS}}}$ does not change, when increasing the spacetime volume by the above mentioned factor $16$. The smaller of the two lattice volumes investigated roughly corresponds to our ensemble at $\beta = 4.35$, which we will use to determine the final result for $\Lambda_{\overline{\textrm{MS}}}$. Therefore, we do not expect considerable finite volume corrections.


\subsubsection{Non-vanishing light quark mass}

Since the perturbative expressions in section~\ref{sec:pertth} are derived under the assumption of massless dynamical quarks, while our lattice results employ light quarks of equal, but finite mass, with pion masses $m_\textrm{PS} \gtapprox 284 \, \textrm{MeV}$, it is essential to investigate the quark mass-dependence. To this end, we consider the three different values of the light quark mass, available at $\beta = 4.05$ (cf. Table~\ref{TAB077}). As before we extract $\Lambda_{\overline{\textrm{MS}}}$ by performing fits with identical parameters $r_\textrm{min} = 2.53 \times a$ and $r_\textrm{max} = 4.42 \times a$:
\begin{itemize}
\item $m_\textrm{PS} = 325 \, \textrm{MeV}$: \quad
$\rightarrow$\quad $\Lambda_{\overline{\textrm{MS}}} = 312(1) \, \textrm{MeV}$.

\item $m_\textrm{PS} = 449 \, \textrm{MeV}$: \quad
$\rightarrow$\quad $\Lambda_{\overline{\textrm{MS}}} = 313(1) \, \textrm{MeV}$.

\item $m_\textrm{PS} = 517 \, \textrm{MeV}$: \quad
$\rightarrow$\quad $\Lambda_{\overline{\textrm{MS}}} = 312(1) \, \textrm{MeV}$.
\end{itemize}
Within tiny statistical errors $\Lambda_{\overline{\textrm{MS}}}$ is constant in the quark mass region investigated. Therefore, we do not expect a dramatic change, when approaching the massless limit. In other words we consider the systematic error introduced by comparing a massive lattice computation with a massless perturbative calculation negligible compared to the uncertainty of $\pm 26 \, \textrm{MeV}$ already determined in section~\ref{SEC329}.


\subsection{Final results for $\Lambda_{\overline{\textrm{MS}}}$}

In the following, we quote our final results for $\Lambda_{\overline{\textrm{MS}}}$ in QCD with $n_f = 2$ dynamical quark flavors. We list these results both in units of $r_0$\footnote{$r_0$ is defined via $r_0^2 F(r_0) = 1.65$, where $F(r) = {\rm d}V(r) / {\rm d}r$ \cite{Sommer:1993ce}.} and in $\textrm{MeV}$. The determination is based on the lattice results at our finest lattice spacing, i.e.\ at $\beta = 4.35$, as explained in section~\ref{SEC444}. The errors, which we combine by adding them in quadrature, are
\begin{itemize}
\item[(1)] the systematic errors associated with perturbation theory and the input parameters of the fitting procedure (cf.\ section~\ref{SEC444}),

\item[(2)] the estimated lattice discretization errors (cf.\ section~\ref{SEC330}),

\item[(3)] the errors associated with $r_0 / a = 9.81(13)$ and the lattice spacing $a = 0.0420(17) \, \textrm{fm}$, respectively.
\end{itemize}
Therewith, we finally obtain
%
%
%
%
\begin{eqnarray}
\label{EQN844} r_0\Lambda_{\overline{\textrm{MS}}}  \ \ = \ \ 0.658(55) \quad , \quad \Lambda_{\overline{\textrm{MS}}} \ \ = \ \ 315(30) \, \textrm{MeV} .
\end{eqnarray}


\newpage

\section{\label{sec:conclusions}Conclusions}

We have determined $\Lambda_{\overline{\textrm{MS}}}$ for QCD with $n_f = 2$ dynamical quark flavors, by fitting various orders of the perturbative expansion of the $Q\bar Q$ static potential, up to terms of ${\cal O}(\alpha_s^4)$ and $\sim\alpha_s^4\ln\alpha_s$, to corresponding lattice data points.

In units of the hadronic scale $r_0$ our result reads
\begin{eqnarray}
r_0 \Lambda_{\overline{\textrm{MS}}} \ \ = \ \ 0.658(55) ,
\end{eqnarray}
while in physical units it is given by
\begin{eqnarray}
\Lambda_{\overline{\textrm{MS}}} \ \ = \ \ 315(30) \, \textrm{MeV} . \label{eq:final}
\end{eqnarray}
To obtain the latter result, the physical scale has been set by fixing the lattice scale via the experimental results for the pion mass and the pion decay constant.\footnote{Since the lattice setup is $n_f = 2$ QCD, while experimental results may also be affected, e.g.\ by dynamical strange and charm quarks, isospin breaking or electromagnetic effects, a different choice of observables, for example the pion mass and the nucleon mass, might yield a slightly different lattice spacing and, therefore, a corresponding change of the result for $\Lambda_{\overline{\textrm{MS}}}$ in units of $\textrm{MeV}$. The effect might be as large as $10 \%$, i.e.\ of the same order of magnitude as the error $\pm 30 \, \textrm{MeV}$ already quoted. Therefore, specifying $\Lambda_{\overline{\textrm{MS}}}$ in units of $r_0$ is preferable. For a more detailed discussion about problems and ambiguities associated with scale setting cf.,\ e.g.\ \cite{Wagner:2011fs}.}

All sources of systematic error have been investigated: neglect of higher orders in the perturbative expansion, the choice of the scale $\mu$, at which $\alpha_s$ is defined, the dependence of the fit results on the fitting range, lattice discretization errors and finite volume effects, as well as finite quark masses. The errors quoted above are dominated by the variations of $\Lambda_{\overline{\textrm{MS}}}$, when fitting different perturbative expressions for the static potential to the lattice results, while varying the fitting window and the scale $\mu$. In principle this error could be reduced. This, however, would require either higher orders in the perturbative expansion, or lattice results at even finer lattice spacing; neither is available at the moment.

\begin{figure}[htb]
\begin{center}
\input{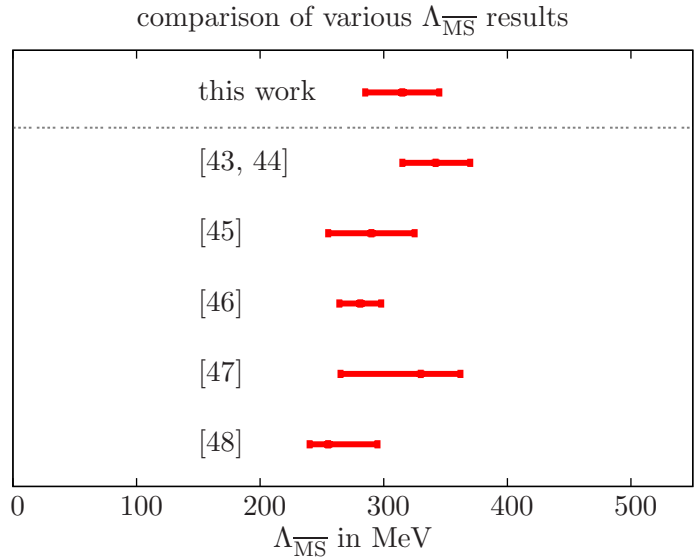}
\caption{\label{FIG005}Graphical summary of various results for $\Lambda_{\overline{\textrm{MS}}}$ in QCD with $n_f=2$ dynamical quark flavors (to eliminate discrepancies due to different methods of scale setting, the results from \cite{DellaMorte:2004bc,Leder:2010kz,Gockeler:2005rv,Sternbeck:2010xu} have been converted using our value for $r_0$, $r_0^\textrm{ETMC} = 0.42 \, \textrm{fm}$).}
\end{center}
\end{figure}

Our results compare well with other determinations of $\Lambda_{\overline{\textrm{MS}}}$ for QCD with $n_f=2$ dynamical quark flavors from the literature, e.g.\
\begin{itemize}
\item a lattice computation using the Schr\"odinger functional \cite{DellaMorte:2004bc,Leder:2010kz}; result: $r_0\Lambda_{\overline{\textrm{MS}}} = 0.73(3)(5)$;

\item a lattice computation using $r_0$ and a boosted coupling \cite{Gockeler:2005rv}; result: $\Lambda_{\overline{\textrm{MS}}} = 261(17)(26) \, \textrm{MeV}$ (the scale has been set via $r_0^\textrm{\cite{Gockeler:2005rv}} = 0.467 \, \textrm{fm}$, while our ETMC $r_0$ is significantly smaller, $r_0^\textrm{ETMC} = 0.42 \, \textrm{fm}$; converting the result to our scale yields $\Lambda_{\overline{\textrm{MS}}} = 290(19)(29) \, \textrm{MeV}$);

\item a lattice computation using Landau-gauge gluon and ghost correlations \cite{Sternbeck:2010xu}; result: $r_0\Lambda_{\overline{\textrm{MS}}} = 0.60(3)(2)$;

\item an ETMC lattice computation using the ghost-gluon running QCD coupling \cite{Blossier:2010ky}; result: $\Lambda_{\overline{\textrm{MS}}} = 330(23)(22)_{-33} \, \textrm{MeV}$;

\item variationally optimized perturbation, combined with renormalization group properties \cite{Kneur:2011vi}; result: $\Lambda_{\overline{\textrm{MS}}} = 255_{-15}^{+40} \, \textrm{MeV}$.
\end{itemize}
Our result for $\Lambda_{\overline{\textrm{MS}}}$ and the results from literature listed above are summarized graphically in Figure~\ref{FIG005}.

Let us finally emphasize again that our value for $\Lambda_{\overline{\textrm{MS}}}$ for QCD with $n_f=2$ is in good agreement with the results of other approaches aiming at its determination (cf. Figure~\ref{FIG005}). The systematic error in our final result, eq.~(\ref{eq:final}), is of the order of ten percent, which is also compatible with the systematic errors of the works 
listed in Figure~\ref{FIG005}. Our investigation here opens the path towards the determination of $\Lambda_{\overline{\textrm{MS}}}$ for a larger number of dynamical quark flavors also, such as the ongoing
efforts of the ETMC collaboration using $N_f=2+1+1$ (dynamical up and down, strange and charm quarks) \cite{Baron:2010bv,Baron:2010th}.
As we were able to show in this work, that $\Lambda_{\overline{\textrm{MS}}}$ can be determined from the static potential for a lattice spacing of $a \approx 0.042 \, \textrm{fm}$, a corresponding calculation for $N_f=2+1+1$ 
will hence be possible as soon as respective lattice data at such small values of the lattice spacing becomes available.


\newpage

\appendix

\section{\label{SEC411}Monte Carlo evaluation of the integral (\ref{EQN612})}

In order to achieve a tree-level improvement of the static potential, computed with the tree-level improved Symanzik action, one has to solve the integral
\begin{eqnarray}
\label{EQN533} G(\mathbf{r}) \ \ = \ \ \frac{1}{(2 \pi)^3} \int_{-\pi}^{+\pi} d^3k \frac{\prod_{j=1}^3 \cos(k_j r_j)}{4 \sum_{j=1}^3 \sin^2(k_j/2) + (4/3) \sum_{j=1}^3 \sin^4(k_j/2)}
\end{eqnarray}
(cf.\ section~\ref{SEC599} and eq.\ (\ref{EQN612})).

In principle one could evaluate this integral directly by means of standard Monte Carlo methods. However, due to the fact that the integral is dominated by the singularity at $\mathbf{k} = 0$, this would result in a rather large statistical error.

To circumvent this problem, we split the integral in two parts,
\begin{eqnarray}
 & & \hspace{-0.7cm} G(\mathbf{r}) \ \ = \ \ G_1(\mathbf{r}) + G_2(\mathbf{r}) \\
 & & \hspace{-0.7cm} G_1(\mathbf{r}) \ \ = \ \ \frac{1}{(2 \pi)^3} \int_{-\pi}^{+\pi} d^3k \frac{\prod_{j=1}^3 \cos(k_j r_j)}{4 \sum_{j=1}^3 \sin^2(k_j/2)} \\
\nonumber & & \hspace{-0.7cm} G_2(\mathbf{r}) \ \ = \ \ \frac{1}{(2 \pi)^3} \int_{-\pi}^{+\pi} d^3k \bigg(\frac{\prod_{j=1}^3 \cos(k_j r_j)}{4 \sum_{j=1}^3 \sin^2(k_j/2) + (4/3) \sum_{j=1}^3 \sin^4(k_j/2)} - \frac{\prod_{j=1}^3 \cos(k_j r_j)}{4 \sum_{j=1}^3 \sin^2(k_j/2)}\bigg) . \\
 & & \hspace{-0.7cm}
\end{eqnarray}
The first part, $G_1$, can be solved exactly by applying a clever recursion technique \cite{Luscher:1995zz}. Since the integrand of the second part is rather smooth, a Monte Carlo evaluation yields a statistical error, which is around two orders of magnitude smaller than that obtained by a direct computation of (\ref{EQN533}), when using a comparable amount of computer time. In other words the necessary computational resources needed to reach a desired precision are reduced by a factor of $\approx 10^4$.




\newpage

\section*{Acknowledgments}

It is a pleasure to thank E.~Garcia Ramos and A.~Nube for simulating and providing ETMC gauge link configurations at $\beta = 4.35$.

F.K.\ is grateful to N.\ Brambilla and A.\ Vairo for drawing his attention to the issue of the determination of $\Lambda_{\overline{\rm MS}}$ by merging perturbative calculations and lattice simulations, and acknowledges detailed and enlightening discussions with them.

Moreover, we would like to thank J.\ Braun, R.\ Sommer and M.\ Steinhauser for helpful discussions.

This work has been supported in part by the DFG Sonderforschungsbereich TR9 Computergest\"utzte The\-o\-re\-tische Teilchenphysik.

M.W.\ acknowledges support by the Emmy Noether Programme of the DFG (German Research Foundation), grant WA 3000/1-1.

F.K.\ and M.W.\ acknowledge inspiring comments from K\"onig M.~Dillig.



\end{document}